\documentclass{llncs}
\usepackage[utf8]{inputenc}
\title{RV17}

\usepackage{wrapfig}
\usepackage{lipsum}
\usepackage{tikz}
\usepackage{xspace}

 \usepackage{algorithm}
\usepackage{algpseudocode}
\usepackage{amsmath}
\usepackage{subcaption}
\usepackage{url}

\newtheorem{defn}{Definition}

\newcommand{\hb}{\xrightarrow{hb}}

\usepackage{verbatim}
\begin{document}
\newcommand{\regionsize}{\ensuremath{\mathcal{RS}}\xspace}
\newcommand{\br}[1]{\ensuremath{\langle #1 \rangle}\xspace}
\newcommand{\maxdep}{\ensuremath{v}\xspace}
    \renewcommand{\topfraction}{0.9}	
    \renewcommand{\bottomfraction}{0.8}	
    \setcounter{topnumber}{2}
    \setcounter{bottomnumber}{2}
    \setcounter{totalnumber}{4}     
    \setcounter{dbltopnumber}{2}    
    \renewcommand{\dbltopfraction}{0.95}	
    \renewcommand{\textfraction}{0.07}	
    \renewcommand{\floatpagefraction}{0.7}	
    \renewcommand{\dblfloatpagefraction}{0.7}	

\title{Monitoring Partially Synchronous Distributed Systems using SMT Solvers}
%
\author{Vidhya Tekken Valapil\inst{1}, Sorrachai Yingchareonthawornchai\inst{1},
Sandeep Kulkarni\inst{1},  Eric Torng\inst{1} \and Murat Demirbas\inst{2}}
%
%
\institute{Michigan State University, East Lansing MI 48823, USA,\\
\email{tekkenva@cse.msu.edu,yingchar@msu.edu,sandeep@cse.msu.edu,torng@msu.edu},\\ 
\and
University of Buffalo,
SUNY Buffalo, Buffalo, NY 14260, USA\\
\email{demirbas@buffalo.edu}}
\maketitle              

\begin{abstract}
In this paper, we discuss the feasibility of monitoring partially synchronous distributed systems to detect latent bugs, i.e., errors caused by concurrency and race conditions among concurrent processes. We present a monitoring framework where we model both system constraints and latent bugs as Satisfiability Modulo Theories (SMT) formulas, and we detect the presence of latent bugs using an SMT solver.
We demonstrate the feasibility of our framework using both synthetic applications where latent bugs occur at any time with random probability and an  application involving exclusive access to a shared resource with a subtle timing bug. 
We illustrate how the time required for verification is affected by parameters such as communication frequency, latency, and clock skew.
Our results show that our framework can be used for  real-life applications, and because our framework uses SMT solvers, the range of appropriate applications will increase as these solvers become more efficient over time.
\end{abstract}

\section{Introduction}
In this paper, we focus on runtime monitoring of latent concurrency bugs in loosely synchronized, distributed, safety-involved systems with the help of SMT solvers. 
By distributed, we mean that the processes/components are linked by a network and communicate with each other by passing messages, though our approach could also be used for shared memory processes. 
By loosely synchronized, we mean that the processes employ some form of clock synchronization such as NTP ensuring that clock drift is bounded by some specified value $\epsilon$. 
By safety-involved, we mean that failure of the distributed system may put lives or the environment at risk; in the extreme, these systems may be safety-critical where failures would lead to loss of life and/or environmental damage.

Examples of such systems include embedded systems, e.g., different components in a car that communicate with each other, vehicular systems, a set of cars on a highway that need to coordinate with each other to avoid collision and maximize their performance, and distributed sensor networks that are used for intrusion detection.

Since we are dealing with safety-involved and potentially safety-critical systems, we must ensure that the deployed systems function correctly.
Unfortunately, due to the inherent uncertainty and complexity of these systems, it is difficult to eliminate all bugs before deployment.
Therefore, it is necessary to monitor deployed systems at runtime to ensure that they function correctly and to detect any violations of safety specifications as early as possible.

One of the most challenging correctness issues of complex distributed systems is latent concurrency bugs that are caused by concurrency issues/race conditions.
Since we assume that the processes are only loosely synchronized, we cannot totally order all the events in the system; but the events obey some partial order.
This means that any observation of the system is some serialization of the partial order of events.
We define latent concurrency bugs to be bugs that are only visible in some but not all serializations of the partial order of events in the system.
To simplify terminology, we refer to latent concurrency bugs as latent bugs in the rest of the paper.

Latent bugs are important, as they indicate a potential for something to go wrong. 
If detected early, these bugs may be fixed by using stronger synchronization, introducing delays and so on.
Unfortunately, identifying latent bugs is a very challenging problem.
Straightforward enumeration of all serializations is not efficient as the number of serializations is likely exponential in the several factors.
In fact, in some cases, identifying latent bugs is an NP-hard problem \cite{PredDetNpComplete}.

We propose to address this challenge using SMT solvers.
An SMT solver takes a formula and a list of constraints as its input.
The solver then identifies whether the formula can be satisfied while simultaneously satisfying all constraints.
If satisfiable, it produces a satisfying variable assignment for the formula.
Otherwise, it reports that the formula cannot be satisfied.
We propose to use SMT solvers as follows.
First, we develop a formula to represent that the violation of a safety specification. This formula is developed once for the system.
Then during runtime, we propose a lightweight method for the monitor to generate the system events and constraints that define the partial order on the system events that any serialization must follow.
The SMT solver then determines if there is a serialization of events that would lead to violation of the safety specification.
Our main focus is on developing the lightweight method for generating system events and constraints that define the partial order on system events.

Relying on SMT solvers for runtime monitoring has several advantages. The most important advantage is correctness. Since an SMT solver evaluates all possible combinations of variables before declaring the formula unsatisfiable, it guarantees the correctness of the monitor; i.e., it will not miss an error and it will not identify phantom errors. Also, the field of SMT solvers is an  active field where new advances result in more efficient solvers. Thus, over time, runtime monitors based on SMT solvers will be able to monitor more complex systems.

We give two justifications for the use of SMT solvers. First, we show that monitoring a distributed system with perfect accuracy, concurrent execution and efficiency (ACE) is \textit{impossible} unless P=NP; this result is a restatement of a known result in asynchronous systems \cite{PredDetNpComplete}.
The second justification is that the impossibility result is for the worst case. In practice, traces often have some structure \cite{EventRateIndepBasin} that can be exploited by highly optimized SMT solvers. The major question regarding the use of SMT solvers in performing runtime monitoring is whether they are fast enough to allow the monitor to keep up with the system processes. This is a valid question since we are asking them to solve potentially NP-hard satisfiability input instances on the fly. We note that any runtime monitoring solution that guarantees correctness has to solve the same problem, so the difficulty of keeping up is not limited to SMT solvers.
With this motivation, we present an algorithm to map runtime execution of distributed programs into instances that can be evaluated using SMT solvers. We use the SMT solver Z3 \cite{Z3} for this purpose. We also analyze the effectiveness of using Z3 in two applications: first in a synthetic application to evaluate the role of different system parameters (communication frequency, clock skew etc). The second in a shared memory access program that has a subtle bug. 

\textbf{Organization of the paper. }  \ 
The rest of the paper is organized as follows. 
In Section \ref{sec:preliminaries}, we define the system, monitor model and introduce the monitoring problem. 
We show how any monitor must choose among accuracy, concurrency, and efficiency in
Section \ref{sec:impossible}. We illustrate latent bugs in Section \ref{sec:latent}. We describe the necessary instrumentation in Section \ref{sec:instrumentation} and how to generate the SMT formulas in Section \ref{sec:generatesmt}. We present our experimental results in Section \ref{sec:exp}. Finally, we discuss related work in Section \ref{sec:related} and provide concluding remarks in Section \ref{sec:concl}.

\section{Preliminaries}
\label{sec:preliminaries}
\subsection{System Model}
\label{subsec:systemmodel}
Our system model is similar to the quasi-synchronous model in \cite{TimedAsychDistrModel}. We consider a system that consists of $n$  application processes numbered $1$ to $n$ where each process $i$ has its own clock. 
We assume that the underlying system guarantees that clocks of any two processes differ by at most $\epsilon$, the clock skew, by using a protocol such as NTP. 
The processes communicate via messages. The minimum and maximum message delays between processes are $\delta_{min}$ (could be $0$) and $\delta_{max}$ (could be $\infty$), respectively.
Each process $i$ is also associated with a single variable $v_i$. 
Our techniques can be easily extended to processes having multiple variables.
Each process execution is a sequence of events.
The two main events are message send or receive events and variable events (the variable changes its value).
The local clock when event $e$ occurred at process $i$ is denoted by $pt.i(e)$.

\subsection{Monitor Model}
\label{subsec:monitormodel}
In our initial discussion, we assume that monitoring for latent bugs is performed by one or more dedicated monitoring processes that are different from the application processes. (During analysis of experimental results, we also consider an alternate implementation where each process devotes a part of its computational resource to the task of monitoring.)
Each application process reports its events to the monitor using messages. 
We assume that the messages from each application process arrive at the monitor in a FIFO order for that process. 

We assume that we can characterize latent bugs with a predicate $P$ that is defined over $n$ variables of the application processes.
For example, if we were implementing a token passing structure, $v_i$ might be a Boolean variable that denotes that process $i$ has the token, and an event $e$ would occur when process $i$ takes the token to change $v_i$ from 0 to 1 and when process $i$ releases the token $v_i$ changes from 1 to 0.
$P$ would be that there is no time $t$ where $\sum_{i=1}^n v_i>1$, i.e., the token is never possessed by more than one process simultaneously.

The monitor processes the events it receives from the application processes to determine if there is a legal serialization of events such that predicate $P$ is true in that serialization.
We evaluate a monitor in terms of precision, recall, and latency.
By precision, we mean that if the monitor declares that predicate $P$ as true, then some legal serialization (defined precisely in Section \ref{sec:concur}) of events will cause the system to reach a state where $P$ is true. 
By recall, we mean that if some legal serialization of application events causes the system to reach a state where $P$ is true, then it is detected by the monitor. 
By latency, we mean the time spent between reaching a state where  $P$ is true and the monitor concluding that  $P$ is true. 
We define a monitor to be a $\Delta$-latency monitor if at any time $t$, the monitor can verify whether $P$ has been satisfied by time $t- \Delta$.
Ideally, we would like to have $0$-latency monitors, but this is not possible for a variety of reasons including message delay and processing time.
Instead, we try to minimize  $\Delta$.

\subsection{Concurrent Events, Happened Before Relation, Valid Snapshots}
\label{sec:concur}
We briefly recall notions of concurrent events, happened before relation and consistent snapshots \cite{HolyGrailSchwarz1994}. 
We define the goal of monitoring as determining if there is some legal serialization of application events that causes the system to reach a state where $P$ is true.
By state, we mean an assignment of values for the $n$ process variables.
We now define what is a legal serialization of application events and how the monitor might detect when the system could reach a state where $P$ is true.
A priori, we assume that all events might be concurrent and thus all serializations are legal.
We rule out some events from being concurrent and thus some possible serializations using  happened before relation, where event $A$ cannot be concurrent with event $B$ if $A$ happened before $B$, or vice versa. 
\begin{defn}
Given two events $A$ and $B$, we say that event $A$ happened before event $B$, denoted as $A \hb B$ iff one of the following four conditions holds.
\begin{itemize} 
\item {\bf Local Events.} Events $A$, $B$ are at the same process $i$ and $pt.i(A) < pt.i(B)$.
\item {\bf Communication.} $A$ is a send event, $B$ is the corresponding receive event.
\item {\bf Clock Synchronization.} Event $A$ happens on process $i$ and event $B$ happens on process $j$ and $pt.j(B) - pt.i(A) > \epsilon$.
\item {\bf Transitivity.} There exists an event $C$ such that $A\hb C$ and $C \hb B$. 
\end{itemize}
\end{defn}
\noindent
Clearly, if $A \hb B$, then in any legal serialization of events, $A$ must appear before $B$.

\begin{defn}
Events $A$, $B$ are possibly concurrent, $A || B$, if $A \not\hb B$ $\land$ $B\not\hb A$.
\end{defn}
\noindent 
If $A || B$, then a legal serialization of events might have $A$ appear before $B$ or $B$ appear before $A$.
If all events are pairwise possibly concurrent, the number of legal serializations of $x$ events would be $|x|!$.
With the partial order defined by the happened before relation, many of these serializations are eliminated.
One of the factors that makes monitoring difficult is if the number of serializations is large.
A common approach for searching for a legal serialization of events is to search for what is known as a 
\textit{consistent snapshot} which we define as follows.
\begin{defn}
A \emph{snapshot} is a set $S$ of $n$ events, one per process. A snapshot $S$ is \emph{consistent} if for any two events $A,B \in S$, $A$ and $B$ are possibly concurrent.
\end{defn}

In our analysis, we assume that frontier events of the snapshot correspond to local events; if the designer wants frontier events to be send events (respectively, receive events) then we create a new local event just before (respectively, after) the event chosen by the designer. 

We need the snapshot to be consistent, and the predicate $P$ to be true in this consistent snapshot.
Thus, we define the following term.
\begin{defn}
A snapshot $S$ is \emph{valid} if and only if
it is consistent 
and the predicate being detected is satisfied at the time of this snapshot.
\end{defn}
Restating the monitoring goal, the monitor strives to find a valid snapshot as soon as possible after that snapshot first exists.

\subsection{Hybrid Logical Clocks}
\label{sec:hlc}
To help the monitor accurately identify when two events might be concurrent or when one event happened before another, we use hybrid logical clocks (HLC) \cite{hlc} to timestamp an event $e$ with an HLC value $hlc.e$.
The local physical time is not sufficient for this purpose because of clock drift.
For example, because of clock drift, the local physical time for a send event $e$ might be larger than the local physical time for the corresponding receive event $f$ even though $e$ clearly happened before $f$.
HLC timestamps provide a simple and efficient way to ensure that if one event $e$ happened before another event $f$, then $hlc.e < hlc.f$.

We now briefly describe how HLC ensures this.
A timestamp $hlc.e$ associated with event $e$ consists of two integers $l.e$ and $c.e$. The value of $l.e$ captures the maximum physical clock value that a process was aware of when event $e$ was created. In many cases, $l.e$ is the same as the physical clock of the process where $e$ was created. However, if this process receives a message with a higher $l$ value than its own clock, $l.e$ reflects that higher value. 
In $hlc.e$, $c.e$ acts as a counter to capture situations where $l.e$ alone cannot determine the timestamp of the newly generated event. 
Also, $hlc.e < hlc.f$ if and only if $(l.e < l.f) \vee ((l.e = l.f) \wedge (c.e < c.f))$.
(Complete algorithm for HLC can be found in \cite{hlc}.)
Since $l.e$ captures the maximum clock that the process was aware of when event e was created, if $|l.e - l.f| < \epsilon$, it is possible that $e$ and $f$ could have happened at the same time. Hence, in the absence of additional information (e.g., a message sent after $e$ and received before $f$), we can treat that they are possibly concurrent. 
Our overall discussion does not depend upon the implementation of HLC; it only relies on its property that it provides logical clocks and that $l.e$ is within $\epsilon$ of the physical clock and that $e \hb f => (l.e < l.f)  \vee ((l.e = l.f) \wedge (c.e < c.f))$.

\section{Worst-Case Impossibility Result}
\label{sec:impossible}
We identify three desirable properties for any monitor:
(1) Accuracy (Precision and Recall), 
(2) Concurrency (Non-intrusiveness), and
(3) Efficiency (Polynomial time execution).
An accurate monitor provides perfect precision and recall which means the monitor claims that the predicate is satisfiable iff there exists a valid snapshot. 
A concurrent or non-intrusive monitor does not interrupt or block the normal execution of the system. 
For example, the monitor never asks a process to delay sending messages or delay performing its computation. 
An efficient monitor performs detection in polynomial time.
We define an ACE monitor to be a monitor that is accurate, concurrent, and efficient.
ACE monitors are desirable 
when runtime verification of distributed system is necessary. 
Unfortunately, Garg's result for asynchronous systems \cite{PredDetNpComplete} also applies to our setting with partially synchronous systems which means that for arbitrary Boolean predicates, ACE monitors are impossible unless P=NP.
In Appendix, we briefly recap Garg's NP-completeness proof highlighting the modification needed to handle our partially synchronous setting.
Although the NP-completeness result implies that there is no general ACE monitor, there is hope for a good monitor in the partially synchronous setting that we study.
First, the NP-hardness reduction requires that all processes have variable events within an $\epsilon$ window.
Second, each application has a specific predicate rather than arbitrary predicates.
Many specific predicates such as conjunctive predicates can be handled in polynomial time.
Also, even for harder predicates, many specific instances may be solved efficiently, especially with modern SAT/SMT solvers.

\section{Latent/Concurrency Bugs}
\label{sec:latent}
We now illustrate latent bugs using a simple protocol for exclusive access to shared data.
To simplify the example, we use only two processes, and the invariant property that we wish to monitor is that at any moment, only one process accesses the shared resource.
Exclusive access can be implemented in many ways such as using time division multiplexing, message passing, or their combination.

We first illustrate how the use of time division multiplexing with improper care for clock skew can lead to a latent bug.
Suppose we use time division multiplexing where Process 1 is given exclusive access to the shared resource in the interval $[0,50)$ while Process 2 is given exclusive access to the shared resource in the interval $[50,100)$.
Further suppose that Process 1 uses its exclusive access in the interval $[45,50)$ while Process 2 uses its exclusive access in the interval $[55,60)$ as shown in Figure  \ref{fig:latentbugs} (a).
If the clock drift $\epsilon < 5$, then this execution is fine and there is no possibility of simultaneous access of the shared resource.
On the other hand, if the clock drift $\epsilon > 5$, then moments $\br{50,0}$ and $\br{55,0}$ are potentially concurrent which means both processes might be simultaneously accessing the shared resource.
In this example, for the given clock drift, process $P_1$ should not access the resource this close to the end of its exclusive access time window to prevent this from occurring.

We next illustrate how messages can potentially ensure proper operation.
Suppose instead of using time division multiplexing, the processes use message passing to pass a token. Suppose Process 1 initially possesses the token exclusively in the interval $[45,50)$, then passes the token to Process 2 in a message that it sends at time $51$ which is received at Process 2 at time $54$  as shown in Figure \ref{fig:latentbugs} (b), and Process 2 exclusively accesses in the interval $[55,60)$.
Because of message $m$, no matter how large the clock skew $\epsilon$ is, moments $\br{50,0}$ and $\br{55,0}$ are not potentially concurrent.
Specifically, $\br{50,0} \hb \br{55,0}$, and thus there is no concurrent access of shared resource.
We will return to this example later to illustrate how we generate the SMT formula necessary for identifying potential errors by performing runtime monitoring using an SMT solver. 

\begin{figure}
\begin{center}
\begin{tikzpicture} [scale=0.55]
\node at (7.5,3.5)  {\textbf{Do both processes possibly simultaneously have a token?}};
\draw [very thin] (0,2.5)--(7,2.5);
\draw [very thin] (0,1)--(7,1);
\node at (0,2.5) [above,left] {P1};
\node at (0,1) [above,left] {P2};
\node at (8,2.5) [above,left] {P1};
\node at (8,1) [above,left] {P2};
\draw (1,2.5)--(1,2.7)--(3,2.7)--(3,2.5);
\draw (4.5,1)--(4.5,1.2)--(6.6,1.2)--(6.6,1);
\node at (0.2,2.2) [below,right]{ \fontsize{8}{10} $\langle45,0\rangle$};
\node at (2,2.2) [below,right]{\fontsize{8}{10}$\langle50,0\rangle$};
\node at (3.8,0.7) [below,right]{\fontsize{8}{10}$\langle55,0\rangle$};
\node at (5.3,0.7) [below,right]{\fontsize{8}{10}$\langle60,0\rangle$};
\node at (3.5,0) {(a) Yes};
\node at (12.6,0.7) [below,right]{\fontsize{8}{10}$\langle 55,0\rangle$};
\node at (14.1,0.7) [below,right]{\fontsize{8}{10}$\langle 60,0\rangle$};
\node at (3.8+7.3,0.7) [below,right]{\fontsize{8}{10}$\langle 54,0\rangle$}; 
\node at (0.3+7.7,2.2) [below,right]{\fontsize{8}{10}$\langle 45,0\rangle$};
\node at (1.8+7.7,2.2) [below,right]{\fontsize{8}{10}$\langle 50,0\rangle$};
\node at (1.8+8.8,2.8) [below,right]{\fontsize{8}{10}$\langle 51,0\rangle$};
\draw [very thin] (0+8,2.5)--(7+8.2,2.5);
\draw [very thin] (8,1)--(7+8.2,1);
\draw (1.5+8-0.5,2.5)--(1.5+8-0.5,2.7)--(2.5+8,2.7)--(2.5+8,2.5);
\draw (4.5+8.7,1)--(4.5+8.7,1.2)--(6+8.9,1.2)--(6+8.9,1);
\draw [->] (11.5,2.5)--(11.9,1.1) node [pos=0.6, auto]{$m$};
\node at (3.5+8,0) {(b) No};
\end{tikzpicture}
\end{center}
\caption{\small{Example of a token passing system with 2 nodes. In  (a), there are four variable events and no messages. Due to clock drift, it is possible that both processes simultaneously share the token if $\epsilon > 5$. Part (b), has the same four variable events plus a message $m$. Because of the message, the two processes cannot share the token regardless of $\epsilon$.}}
\label{fig:latentbugs}
\end{figure}

\section{Instrumentation for Runtime Monitoring}
\label{sec:instrumentation}
In this section, we identify the instrumentation required to support runtime monitoring.
We describe how each application process reports changes in variable values and inter-process messages to the monitor process. In practice if messages are being sent over a network, it could be observed by the monitor directly.  
\subsection{Reporting a change in variable value}
\label{subsec:reportChangeInValue}
Every time when the value of the variable $v_i$ changes, process $P_i$ sends a message with three pieces of information to the monitor: the previous value of $v_i$, the HLC timestamp of the previous variable event, and the current HLC timestamp associated with the new variable event.
The two timestamps are sent as an interval that includes the left endpoint but excludes the right endpoint.

To make this work, we assume process $P_i$ starts with an HLC value of $\br{0,0}$ and initially $v_i = a_i$.
The information for the new variable event 
will be captured in the next variable event message sent to the monitor.
Providing the previous value and HLC timestamp allows the monitor to process messages correctly even if they arrive out of order, though out of order messages may delay detection of predicate satisfaction.
To illustrate these variable event messages, consider the run of the program in Figure \ref{fig:latentbugs} (a) or (b) where each process's Boolean variable $v_i$ is true when process $P_i$ accesses the shared data and is false otherwise.
Process $P_1$ sends two variable event messages to the monitor.
The first message has $v_1 = False$, $[\br{0,0}, \br{45,0})$ and is sent at $\br{45,0}$.
The second message has $v_1 = True$, $[\br{45,0}, \br{50,0})$ and is sent at $\br{50,0}$.
Likewise process $P_2$ sends two messages to the monitor.
The first message has $v_2 = False$, $[\br{0,0}, \br{55,0})$.
The second message has $v_2 = True$, $[\br{55,0}, \br{60,0})$.

\subsection{Reporting Message Events}
\label{subsec:report_msg}
We report inter-process message events by having the process that receives a message report both the send and receive events to the monitor.
Specifically, the process reports four things to the monitor:
the sender process ID and the HLC timestamp for the send event (information that is included in the message by the sender process before sending the message), the receiver process ID, and the HLC timestamp for the receive event.
For example, in  Figure \ref{fig:latentbugs} (b), process $P_2$ sends a message to the monitor with the sender ID $P_1$, the send event timestamp $\br{51,0}$, the receiver process ID $P_2$, and the receive event timestamp $\br{54,0}$.

\section{Generating the SMT Formula}
\label{sec:generatesmt}
Now we illustrate how the monitor will generate a correct formula to send to the SMT solver to detect predicate satisfaction. 
The basic setting is that for each process $P_i$, we have three variables: $v_i$, $l_i$, and $c_i$ that correspond to a variable value and an HLC timestamp for that variable value.
The formula we create will be satisfiable if there is a way to set all $3n$ variables such that the formula is satisfied.
The intuition behind a satisfying variable assignment is that they specify a valid snapshot; i.e., a consistent snapshot where the formula is satisfied.
We add several constraints to ensure that only consistent snapshots will satisfy the SMT formula, some of which are \textit{static constraints} that do not depend on the actual run. Others are  \textit{dynamic constraints} that depend on the actual run.

\textbf{Clock Synchronization Constraints.}
We first enforce the clock synchronization requirement of a consistent snapshot.
Specifically, all the logical clock values $l_i$ must be at most $\epsilon$ apart from each other. We enforce this by adding the following static constraint: 

\indent\indent\indent\indent\indent\indent\indent$\forall i,j \quad 1 \leq i < j \leq n: \vert l_i - l_j \vert \leq \epsilon$

\textbf{Communication Constraints.}
We next enforce all communication requirements of a consistent snapshot.
Specifically, if process $P_i$ sends a message at time $\br{l_s,c_s}$ to process $P_j$ which receives the message at time $\br{l_r, c_r}$, then if process $P_j$'s timestamp in the consistent snapshot is at least $\br{l_r,c_r}$ which means process $P_j$ has received the message, then $P_i$'s timestamp in the consistent snapshot is greater than $\br{l_s,c_s}$ which means that $P_i$ has sent the timestamp.
Thus, for each message reported to the monitor, the monitor adds the following constraint:

\indent\indent\indent\indent\indent\indent$(\br{l_j,c_j} \geq \br{l_r,c_r}) \Rightarrow (\br{l_i,c_i} > \br{l_s,c_s})$

These are dynamic constraints as we need one for every inter-process message.
Continuing with the example discussed in Figure \ref{fig:latentbugs} (b), when the monitor receives the details of message $m$ from process $P_2$, it adds the following constraint:

\indent\indent\indent\indent\indent\indent$(\br{l_2,c_2} \geq \br{54,0}) \Rightarrow (\br{l_1,c_1} > \br{51,0})$

\textbf{Variable Event Constraints.}
We now add constraints to ensure that variable $v_i$ takes on the correct value for consistent snapshot.
We ensure this by adding one constraint per variable event message received by the monitor.
Specifically, if process $P_i$ sends a variable event message $v_i = val$, $[\br{l_1,c_1}, \br{l_2,c_2})$, then we add the constraint:

\indent\indent\indent\indent$(\br{l_i, c_i} \geq \br{l_1, c_1}) \wedge 
 (\br{l_i, c_i} < \br{l_2, c_2}) \ \ \ \Rightarrow \ \ \ v_i = val $

\textbf{Predicate Constraints.}
Finally, we need to ensure that the predicate being monitored is satisfied at the consistent snapshot.
This is a static formula that depends only on the $n$ $v_i$ variables.
For example, if the predicate being monitored requires that all values of $v_i$ are true simultaneously, then it would be captured by adding $\bigwedge v_i$. If the goal is to check that the sum of all $v_i$ values is at least 10, then it would be captured by adding $\sum v_i \geq 10$. 

\subsection{Optimizing By Combining $l$ and $c$ variables}
We now present an optimization where we combine variables $l_i$ and $c_i$ for each process $P_i$ into a new variable $nl_i$ thus eliminating $n$ variables from the formula to speed up the SMT solver.
The basic idea is that the maximum $c$ value in a typical run is very small \cite{hlc}.
Mostly, we do not need the $c$ value.
It is needed to deal with messages that \textit{appear to be from future} due to clock skew.
For e.g., if a process with physical clock $10$ receives a message from a process with $l$ value $20$, the $l$ value of the receive event is set to 20.
To ensure this receive event is later than the send event, the $c$ value of the receive event is set to be one larger than $c$ value of the send event.
This $c$ value will increase if  more events take place at the current process before its physical clock value reaches $l$.
For small $\epsilon$, this is relatively unlikely to happen.
Once $l$ is reset, typically $c$ value will return to 0.

Let $c_{max}$ denote the largest $c$ value encountered during the run.
Let $c' = c_{max}+1$.
We can combine $l_i$ and $c_i$ by creating a new variable $nl_i = c' l_i + c_i$.
That is, the monitor still receives HLC timestamps with $l_i$ and $c_i$ values.
The monitor combines them into $nl_i$ values before sending them to the SMT solver.
Specifically, we modify the constraints as follows.
Previous work showed that $c_{max} \le 3$ for typical parameter values;  we typically used $c' = 4$ in our experiments.
The clock synchronization constraints change from 
$\forall i,j \quad 1 \leq i < j \leq n: \vert l_i - l_j \vert \leq \epsilon$ into
$\forall i,j \quad 1 \leq i < j \leq n: \vert nl_i - nl_j \vert \leq c' \epsilon$.

Likewise, the communication constraints change from
$(\br{l_j,c_j} \geq \br{l_r,c_r}) \Rightarrow (\br{l_i,c_i} > \br{l_s,c_s})$ into
$(nl_j \geq c'l_r+c_r) \Rightarrow (nl_i > c'l_s + c_s)$.
And, the variable event constraints change from
$(\br{l_i, c_i} \geq \br{l_1, c_1}) \wedge 
 (\br{l_i, c_i} < \br{l_2, c_2}) \ \ \ \Rightarrow \ \ \ v_i = val $
 into
 $ (nl_i \geq c'l_1+c_1)  \wedge 
 (nl_i < c'l_2 + c_2) \ \ \ \Rightarrow \ \ \ v_i = val $. Finally, no changes are needed for the predicate constraints since they do not use HLC timestamps.

\section{Experimental Results}
\label{sec:exp}
We now present our  experimental results. 
We use a system of 10 independent processes where their clocks differ by at most $\epsilon$. When a process is running, it sends messages to randomly selected processes at some communication frequency $mfr$. Each message is received after time $\delta$. 
In one set of experiments, we use a synthetically generated workload where process variable $v_i$ changes value randomly; we consider $v_i$ as both Boolean and integer variables.
In another set of experiments, we use an exclusive access to the shared resource in a shared-resource application that has a timing error that can potentially cause two processes to simultaneously access the shared memory.
We run our experiments for one second of actual time where we generate event messages  as described in Section \ref{sec:instrumentation}. The monitor generates SMT constraints as described in Section \ref{sec:generatesmt}.
We then run Z3 on the SMT formula.
In our simulation, SMT is invoked periodically (period chosen to be 1s). It could also be changed so that it is invoked when a new event is received (or when a given threshold number of events is received). 

In our experiments, our default parameters are a clock tick of $0.01ms$, a clock drift $\epsilon = 10 ms$ (1000 clock ticks), message delay $\delta = 1 ms$ (100 clock ticks), $\beta = 1\%$ (the expected time before the variable becomes true is $1ms$), $interval = 0.1 ms$ (10 clock ticks) and an average communication frequency of 1000 messages per second (1\% chance of sending a message every clock tick).
Among all the experiments performed, the predicate of interest is satisfiable approximately 70\% of the time. Since we avoid generating instances where the satisfaction of the predicate of interest is \textit{too easy}, we do not observe a clear pattern that indicates a correlation between the time taken by Z3 and whether the predicate of interest is satisfiable, so we omit discussion of whether the given predicate is satisfiable.
However, the raw data from the experiments is available at \url{http://cse.msu.edu/~tekkenva/z3monitoringresults/}.

\textbf{Synthetic workload}
In our synthetic workload, the $v_i$ variables are either boolean variables or integer variables. When they are integer variables, we restrict them to $\{0,1\}$.
In both cases, whenever $v_i$ is eligible to change, process $i$ changes $v_i$'s value with probability $\beta$. Once $v_i$ changes value, it keeps that value for a minimum length of time $interval$ before becoming eligible to change again.
When $v_i$ is a Boolean variable, we consider three different predicates: the conjunctive predicate  $\bigwedge v_i$ that requires all $v_i$ variables to be true simultaneously, the exactly $5$ predicate, $|\{v_i = true \} |=5$, that requires exactly $5$ $v_i$ variables to be true simultaneously, and the at least $5$ predicate, $|\{v_i = true \} |>=5$, that requires at least $5$ $v_i$ variables to be true simultaneously.
When $v_i$ is an integer variable, we consider two predicates $\Sigma v_i = 5$ and $\Sigma v_i \geq 5$ that are equivalent to the exactly 5 and at least 5 Boolean predicates, respectively.

\textbf{Exclusive access workload.}
We use a time division multiplexing protocol where each process accesses the shared data in its time slot which has length $100ms$ and that the clock drift is at most $10ms$. 
We assume that each process will access the data at the start of its time slot.
To ensure that there is no simultaneous access, each process must stop access $10ms$ before the end of its time slot.
For example, process 1 should access the data in the interval $[0ms,90ms)$, process 2 should access it in the interval $[100ms,190ms)$, and so on.
We introduce a chance of error where each process holds on to its access for an extra $1ms$ with a probability of 10\% which means process $i$ and process $i+1$ might simultaneously access the data.
For this experiment, $v_i$ is a Boolean variable that marks when process $i$ is accessing the shared data, and the predicate is whether two $v_i$ variables might be simultaneously true.

In the rest of this section, we first describe how we discretize time. Then, we identify how one can interpret the results of our experiments. Finally, we present the effect of communication frequency, communication latency,  variable stability, and clock skew on the time for monitoring. 

\subsection{Effect of Discretization of Time}
Although time is continuous, we must discretize time to use hybrid logical clocks and SMT solvers.
A natural question is whether the level of discretization  has any affect on the accuracy and efficiency of monitoring.
We first observe that using a clock tick that is too large can have negative modeling effects.
To illustrate this issue, consider the following scenario: clock drift is $10ms$, message delay is $1ms$ and the expected number of messages sent by a process in $1ms$ is one. 
If we model such a system using a clock tick of $1ms$, then the discrete clocks of processes differ by at most $10$ ticks, 
each message is received at the next clock tick, and each process would have to send one message at every clock tick.  In contrast, if use a clock tick of $0.1ms$, then the discrete clocks of processes could differ by at most $100$ ticks, a message would be received after $10$ clock ticks, and a process would send a message with probability $0.1$ at every clock tick and achieve the desired goal of one expected message sent by a process in $1ms$. This is a better model as it allows a process to possibly send no messages or to send multiple messages within $1ms$.
We find that the level of discretization does not have a significant impact on the time required for Z3. Hence, in our analysis, we assume that each clock tick is 0.01ms. In other words, if $\epsilon=10ms$, it would be modeled as $\epsilon=10/0.01=1000$.

\subsection{Interpreting the Experimental Results}
There are two approaches for implementing run-time monitors; a standalone approach where a monitor process is independent of the application process, which is how we have described the monitor process so far, and a combined approach where the monitor runs on the same machines as the application processes and uses a certain fraction of resources from those machines. 
We now describe how to interpret our Z3 timing results using these two perspectives.
Recall that we run the application process for one second in all experiments.

Let us start with the standalone monitor.
If the monitoring time is at most one second, a single monitor running on the given environment (Windows 8.1 on 2.19 GHz Intel(R) Core(TM) i5 and 8.00 GB RAM) would suffice with a latency of at most $1s$. If the monitoring time is more than one second, say two seconds, then we need two machines and two instances of Z3.
If two monitors are used then it could be achieved by sending events at odd time (first, third, fifth second) being sent to the first monitor and sending events at other times to the second monitor. Some overlap may be necessary to ensure that events that span across boundary are recorded correctly.
In general, if the Z3 monitoring time (time required for solving the SMT problem) is $c$ seconds, then we need $\lceil c \rceil$ machines and $\lceil c \rceil$ instances of Z3 to keep pace, and the latency would increase to $c$ seconds.
Note that we can reduce the machine requirements and latency by getting a more efficient machine or finding a more efficient SMT solver.

Let us now consider the combined approach.
In this case, if monitoring one second of execution time on 10 processors takes $c$ seconds, then each process would need to devote roughly $c \times 10\%$ of its resources to the monitor to ensure that the monitoring process can keep up with the application.
We can view this as either needing a $c \times 10\%$ more efficient machine or that
$\frac{c \times 10}{100 + c \times 10}$ of its resources are devoted to monitoring meaning that the application itself will slow down due to monitoring.
The latency in this case will be ten seconds or, in general, $n$ seconds, where $n$ is the number of processors.

\begin{figure}
\subcaptionbox{Effect of communication frequency
}
{\includegraphics[width=0.50\textwidth,height=34mm]{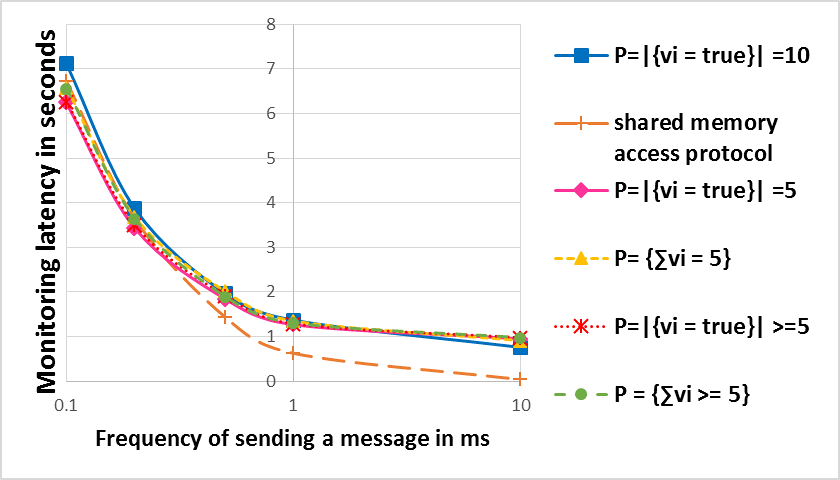}}
\subcaptionbox{Effect of change in message delay $\delta$}
{\includegraphics[width=0.50\textwidth,height=34mm]{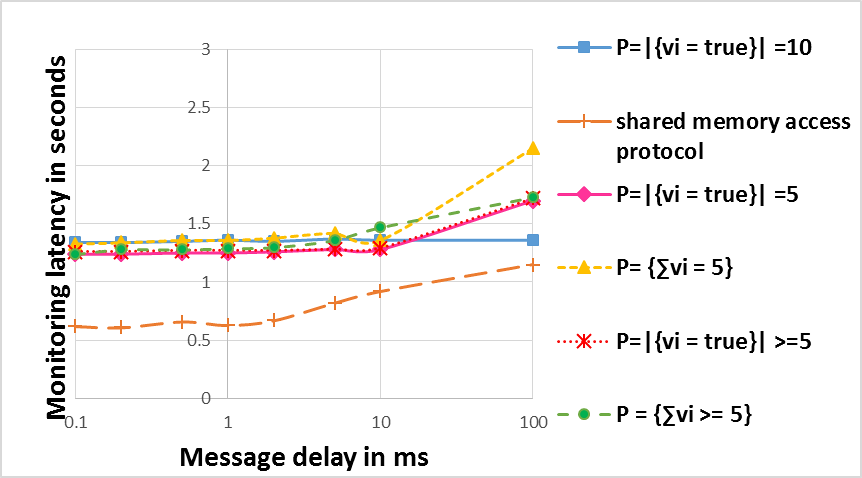}}\vspace{0.1cm}\\
\subcaptionbox{Effect of rate at which the local predicate changes}
{\includegraphics[width=0.5\textwidth,height=33mm]{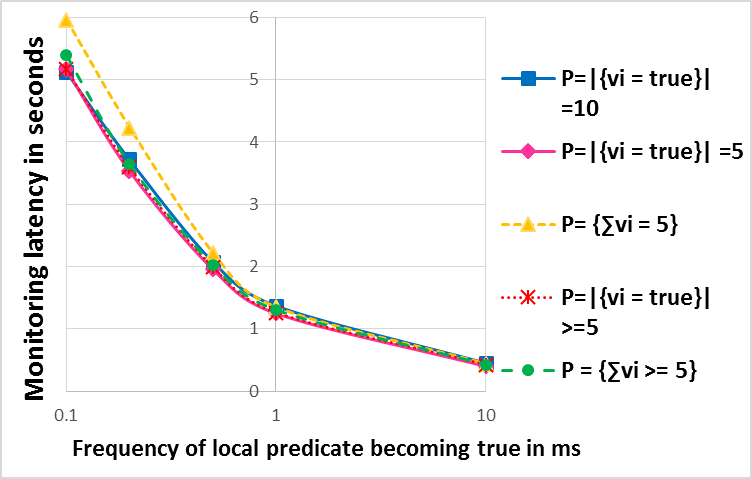}}
\subcaptionbox{Effect of duration for which the local predicate stays unchanged}
{\includegraphics[width=0.5\textwidth,height=33mm]{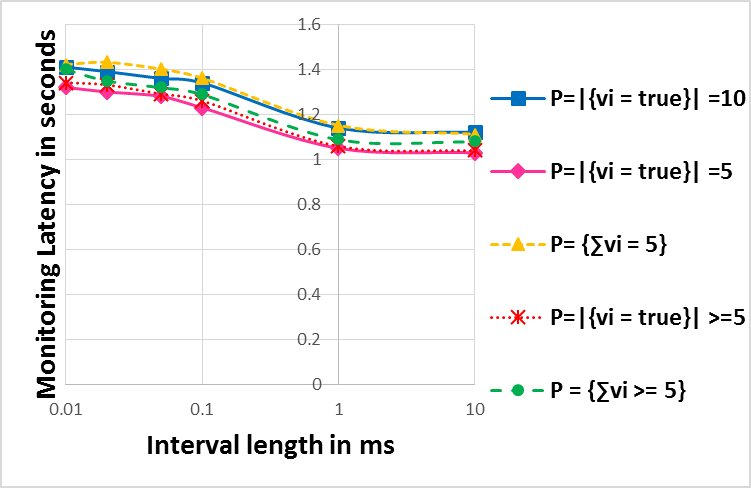}}\vspace{0.1cm}\\
\subcaptionbox{Effect of change in clock drift $\epsilon$}
{\hspace{2cm}
\includegraphics[width=0.6\textwidth,height=33mm]{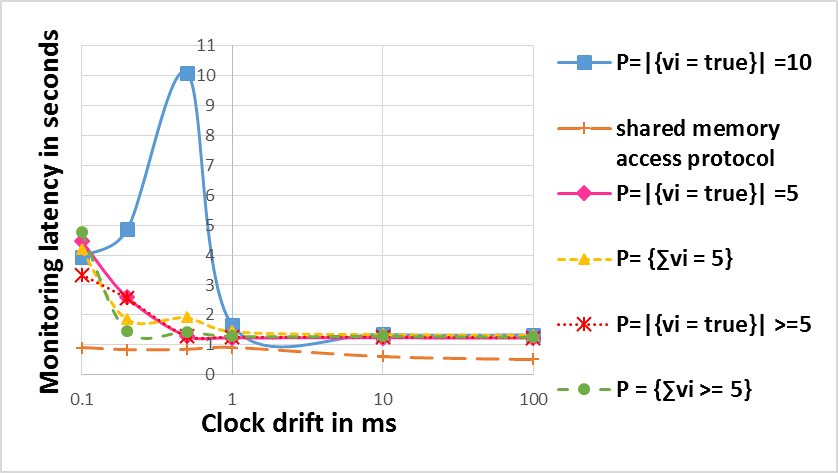}%
}\\
\caption{Analysis of role of system parameters on monitoring latency}
\label{fig:effectofalpha_delta_beta_intervallength_epsilon}
\end{figure}

\subsection{Effect of Communication Frequency}
We first show how inter-process communication frequency affects the time required for monitoring.
Figure \ref{fig:effectofalpha_delta_beta_intervallength_epsilon} (a) summarizes these results. 
We use our default parameters except that we vary communication frequency from an average of 100 messages per second (0.1\% chance of sending a message every clock tick) to an average of 10,000 messages per second (10\% chance of sending a message every clock tick). 
We see that as the communication frequency decreases, the time for verification also decreases. This holds for all predicates we study. 
Also, monitoring the faulty shared memory access protocol requires less time than monitoring the synthetic workloads. 

\subsection{Effect of Communication Latency}
We now show how inter-process communication latency affects the time required for monitoring.
Figure \ref{fig:effectofalpha_delta_beta_intervallength_epsilon} (b) summarizes these results. 
We use our default parameters except that we vary communication latency from  $0.1ms$ to $100ms$. 
We see that communication latency has a small effect on the time required for monitoring.
For all predicates considered, the monitoring time increases with an increase in communication latency, but by at most half a second even when the latency increases from $0.1ms$ to $100ms$.

\subsection{Effect of Variable Stability}
We now show how variable stability affects the time required for monitoring.
Note that there are two parameters that affect variable stability in the synthetic workload experiments: $\beta$ which is the probability of changing the variable value at a given time and $interval$ which determines how long the variable value will remain stable after a change.
We use our default parameters except we first vary $\beta$ from 
$0.1\%$ (the expected time before the variable becomes true is $10ms$) to 
$10\%$ (the expected time before the variable becomes true is $0.1ms$) in one set of experiments and we vary 
$interval$ from
$0.01 ms$ (1 clock tick) to $10ms$ (1000 clock ticks).
Figure \ref{fig:effectofalpha_delta_beta_intervallength_epsilon} (c) summarizes the results where we vary $\beta$ and
Figure \ref{fig:effectofalpha_delta_beta_intervallength_epsilon} (d) summarizes the results where we vary $interval$. 
We see that more variable stability leads to faster monitoring.
As we decrease the probability of changing variable value or increase the stable interval time, Z3 monitoring time drops.

\subsection{Effect of Clock Drift}
We now show how clock drift $\epsilon$ affects the time required for monitoring.
Fig. \ref{fig:effectofalpha_delta_beta_intervallength_epsilon} (e) summarizes these results.
We use our default parameters except that we vary clock drift $\epsilon$ from  $0.1ms$ to $100ms$. 
We see that unlike other parameters, clock drift does not have a monotonic affect on monitoring time.
For some predicates such as  conjunctive predicates, the time for monitoring first increases as $\epsilon$ increases and then decreases as $\epsilon$ increases further.
While we do not know the exact reason for this, we suspect the following is true.
We are looking for a consistent snapshot where the given predicate is true which in some sense requires examining $\epsilon$-length intervals in the execution.
The number of $\epsilon$-length windows is inversely proportional to $\epsilon$.
The number of events within an $\epsilon$ length window and thus the complexity of the window is proportional to $\epsilon$.
Thus, there are competing pressures making the exact complexity a complicated function of $\epsilon$.

\section{Related Work}
\label{sec:related}
\textbf{Distributed Predicate Detection with Vector Clocks.} The fundamental challenge in distributed predicate detection lies in causality induced by inherent non-determinism \cite{HolyGrailSchwarz1994}. 
Most existing distributed system monitoring works have focused on asynchronous systems that use vector clocks (VCs) \cite{fidgeVC,matternVC} and \cite{garg_weak,Chauhan:2013:DAA:2553409.2553427,7056433,PredDetNpComplete,vcsize} make minimal assumptions about the underlying system. Unfortunately, asynchronous monitors have several sources of inefficiency that limit their scalability and impede their adoption in real systems. First, general predicate detection with asynchronous monitors is  NP-complete~\cite{PredDetNpComplete}, although polynomial time algorithms exists for special cases of predicate detection such as linear predicates and conjunctive predicates~\cite{PredDetNpComplete,garg_weak}. Second, asynchronous monitors require $\Theta(n)$ space VC timestamps to track all causalities \cite{vcsize} in the system where $n$ is the number of processes. Thus, each message has linear size in the system which limits scalability of asynchronous monitors.

\noindent
\textbf{Distributed Predicate Detection with Physical Clock.} One way to avoid the overhead of VCs is to use physical time, which has $O(1)$ size timestamps, along with a clock synchronization protocol such as NTP \cite{NTP}, which guarantees that two clocks differ by at most some value $\epsilon$. 
Stoller \cite{Stoller2000} has shown that if  the inter-event spacing is larger than $\epsilon$, then the number of global states that the system can pass through is $O(kn)$ where $k$ is the maximum number of events in any local process; in contrast, in an asychronous system, the number of possible states is $\Omega(k^n)$.
However, the physical time approach fails to rule out many possible interleavings because it ignores the causality implications of messages. Marzullo \cite{Marzullo1992} described another global state enumeration method that included both logical time and physical time with Hybrid Vector Clocks (HVC).
Because of some synchronization, HVC timestamp size may be less than $n$ while preserving properties of VC \cite{HVC}.
In partial synchrony, the trade-off in terms of precision and recall monitoring was discussed in \cite{PrecRecSensOfMonitoring}. Namely, Hybrid Logical Clocks (HLC) \cite{hlc} have been  used for efficient predicate detection when imperfect recall is acceptable \cite{HLCPredicateDetectionICNDCN2017}.

\noindent\textbf{Distributed Runtime Monitoring Beyond Predicate Detection.}
The ultimate goal of runtime monitoring is to monitor expressive properties such as Linear-temporal-logic (LTL) \cite{Sen:2004:EDM:998675.999446,Mostafa:2015:DRV:2863692.2863960}. 
Efficient distributed predicate detection is needed before we can perform efficient distributed LTL monitoring because  LTL formula requires predicate detection as a subtask (e.g., consider P leads to Q). 

Previous work in distributed monitoring of more expressive properties has focused on developing distributed semantics of the centralized counterpart. Gul \cite{Sen:2004:EDM:998675.999446} considered Past Time Distributed Temporal Logic  to express safety  properties  of  distributed  message  passing  systems using vector clocks. Mostafa and Bonakdarpour \cite{Mostafa:2015:DRV:2863692.2863960} give a decentralized monitor for LTL specifications in distributed systems; they focus on sound semantics rather than efficiency. In  \cite{Falcone2014,Bauer:2016:DLM:2979753.2979759}, the authors designed a decentralized approach for monitoring LTL and regular language specifications using an enhanced automaton assuming that global time is accessible to the local processes.  Accessing unambiguous global time requires the use of extremely high precision clocks such as atomic clocks. 

\noindent
\textbf{Monitoring Distributed Systems in Practice.}  Intrusive distributed monitoring is a common choice in practice. For example, Facebook TAO \cite{180185} designed a distributed database by waiting out the $\epsilon$ uncertainty bound during commit-phase so that all events are totally ordered. Google Spanner uses highly synchronized clocks called TrueTime \cite{180268,beyond} which requires transactions to not overlap during the $\epsilon$ uncertainty interval. These systems are relatively easy to monitor since blocking execution reduces explosion of number of concurrent states. 

\section{Conclusion and Future Work}
\label{sec:concl}
In this paper, we focused on the problem of runtime monitoring partially synchronous distributed systems with the help of SMT solver Z3. We showed how one can map the requirements of runtime monitoring into constraints that need to be satisfied. Based on this analysis, we find that the effort for monitoring reduces with a decrease in communication frequency but increases with communication latency. The time for monitoring also decreases when variables involved in the program change less frequently. 

We evaluated our approach for synthetic workload as well as for predicates associated with a program that requires mutual exclusion for shared resource among multiple processes. An interesting observation was that the monitoring time for the synthetic workload was higher.One possible reason is that the constraints created by the synthetic workload does not have any patterns that can be used by the SMT solver. We believe the synthetic workload may represent a hard case and that monitoring may perform better on real protocols. As future work, we plan to test our framework with more predicates and to provide a working tool that can be combined with any application.
\newpage
\bibliographystyle{plain}
\bibliography{sandeep,vidhya,sorrachai-refs}
\appendix

\newpage

\section{Adapting Garg's Proof to Partially Synchronous Systems}
To map Garg's NP-completeness result for asynchronous systems to partially synchronous systems, we create an execution instance as follows.
First, there are no messages.
Second, each process's Boolean variable is initialized to false.
Third, each process has one event that occurs before time $\epsilon$ where the local variable becomes true.
This is where we modify the proof. In the asynchronous setting, the one event where the local variable becomes true can happen at any time.
Since there is no communication and each variable is both true and false within the time interval $[0,\epsilon]$, there exists a legal serialization that produces a state for each of the $2^n$ possible truth assignments for the $n$ Boolean variables.
Thus, there exists a valid snapshot iff the Boolean predicate is satisfiable.
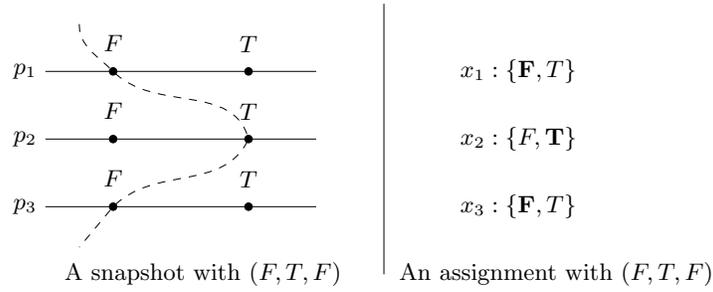
\begin{figure} [ht]
\begin{center}
\begin{tikzpicture} [scale=0.9] 
 \node at (0,1) [left]{$p_3$};  \node at (6,1) [above,right]{$x_3 : \{{\bf F},T \}$};
 \node at (0,2) [left]{$p_2$};\node at (6,2) [above,right]{$x_2 : \{F,{\bf T} \}$};
 \node at (0,3) [left]{$p_1$};\node at (6,3) [above,right]{$x_1 : \{{\bf F},T \}$};
 \node at (4.5,0) [below,left]{A snapshot with $(F,T,F)$}; 
 \node at (5.1,0) [below,right]{An assignment with $(F,T,F)$}; 
\tikzset{every node/.style={draw,shape=circle}}
  \draw [very thin, dashed] (0.5,3.7) to [out=270,in=135] (1,3) to [out=315,in=105] (3,2) to [out=245,in=45] (1,1)--(0.5,0.4);
 \draw  [very thin] (0,3)--(1,3)node[circle,fill,inner sep=1pt,label=above:$F$]{}--(3,3)node[circle,fill,inner sep=1pt,label=above:$T$]{}--(4,3);
   
 \draw  [very thin] (0,2)--(1,2)node[circle,fill,inner sep=1pt,label=above:$F$]{}--(3,2)node[circle,fill,inner sep=1pt,label=above:$T$]{}--(4,2); 
 
  \draw  [very thin] (0,1)--(1,1)node[circle,fill,inner sep=1pt,label=above:$F$]{}--(3,1)node[circle,fill,inner sep=1pt,label=above:$T$]{}--(4,1); 
  
 \draw  [very thin] (5,0)--(5,4);
\end{tikzpicture}
\end{center}
\caption{Example of bijective mapping between a snapshot and corresponding boolean assignment.}
\label{fig:snapshotboolean}
\end{figure}

\end{document}